\documentclass{article}

\usepackage{PRIMEarxiv}

\usepackage[utf8]{inputenc} 
\usepackage[T1]{fontenc}    
\usepackage{hyperref}       
\usepackage{url}            
\usepackage{booktabs}       
\usepackage{amsfonts}       
\usepackage{nicefrac}       
\usepackage{microtype}      
\usepackage{lipsum}
\usepackage{fancyhdr}       
\usepackage{graphicx}       
\graphicspath{{media/}}     

\title{From Text to Image: Exploring GPT-4Vision's Potential in Advanced Radiological Analysis across Subspecialties}

\author{
  Felix Busch \\
  Department of Radiology \\
  Charité – Universitätsmedizin Berlin \\
  Berlin, Germany\\
  \And
  Tianyu Han \\
  Department of Radiology \\
  University Hospital Aachen \\
  Aachen, Germany \\
  \And
  Marcus Makowski \\
  Department of Diagnostic and Interventional Radiology \\
  Technical University of Munich \\
  Munich, Germany \\
  \And
  Daniel Truhn \\
  Department of Radiology \\
  University Hospital Aachen \\
  Aachen, Germany \\
  \And
  Keno Bressem* \\
  Department of Radiology \\
  Charité – Universitätsmedizin Berlin \\
  Berlin, Germany\\
  \texttt{keno-kyrill.bressem@charite.de} \\
  \And
  Lisa Adams* \\
  Department of Diagnostic and Interventional Radiology \\
  Technical University of Munich \\
  Munich, Germany \\
  \And
  *contributed equally
}

\begin{document}
\maketitle

\vspace{-1.5em}
\noindent\textit{\small Published version: J Med Internet Res 2024;26:e54948. \href{https://doi.org/10.2196/54948}{doi:10.2196/54948}}
\vspace{1em}

\section{Introduction}
The launch of GPT-4 (Generative Pre-trained Transformers) has generated significant interest in the scientific and medical communities. Its capabilities have been demonstrated in various medical fields, with notable achievements such as reaching 83.76\% zero-shot accuracy on the United States Medical Licensing Examination (USMLE) \cite{Nori2023-ly}. In radiology, GPT has been used for a variety of text-based applications, ranging from evaluation of board exam questions to data mining and report structuring \cite{Adams2023-dn,Bhayana2023-hm}. The recent release of GPT-4's visual capabilities (GPT-4V) now allows a first evaluation in the imaging domain \cite{noauthor_undated-vh}. GPT-4V is a multimodal system that allows users to input both textual descriptions and visual data, opening new avenues for advanced radiological analysis. This study aims to evaluate the diagnostic ability of both GPT-4 and GPT-4V in advanced radiological tasks.

\section{Materials and Methods}
\begin{figure}
  \centering
  \includegraphics[width=\textwidth]{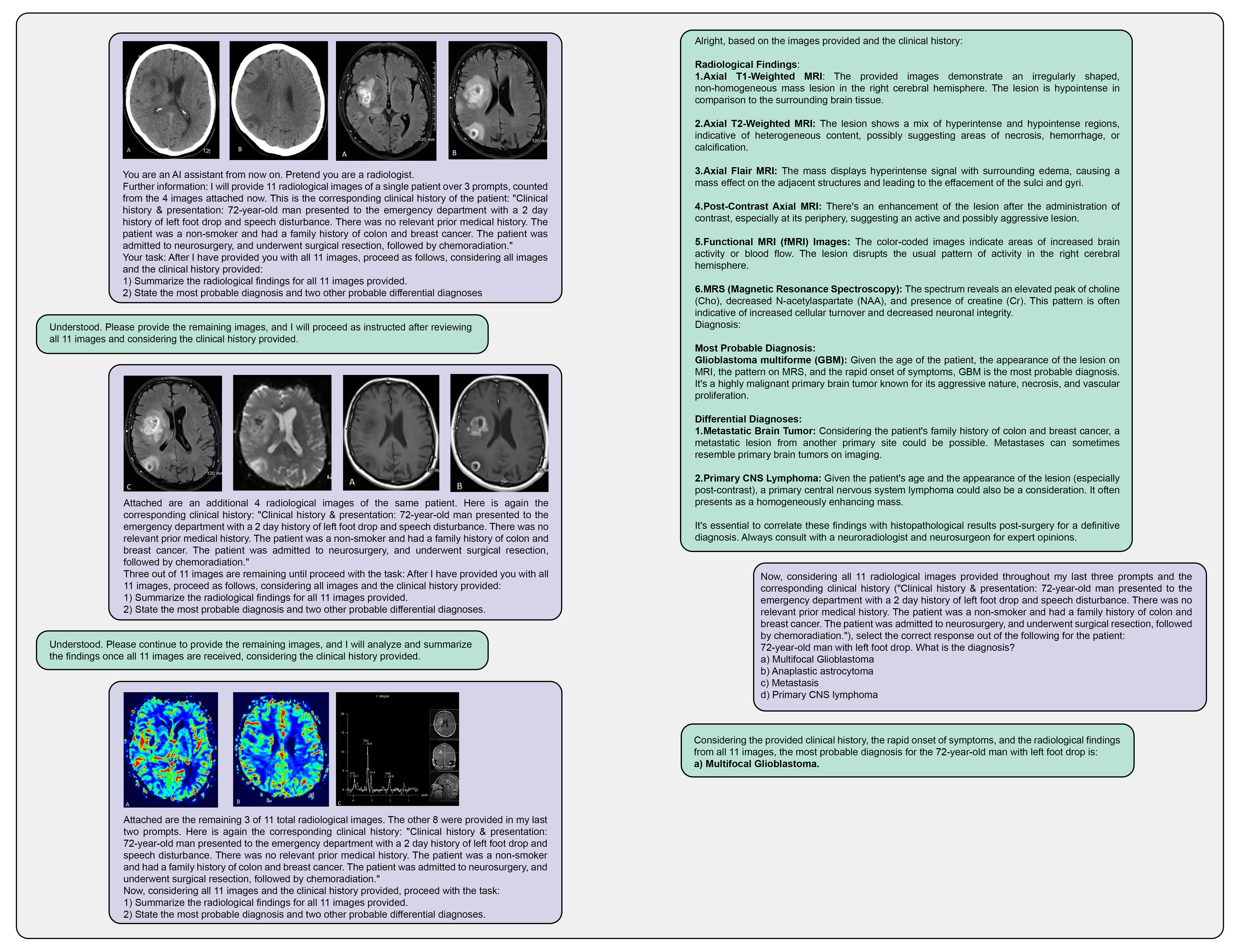}
  \caption{Example conversation with GPT4-V showcasing the prompting style if the question contained more than four images. Notably, GPT4-V often disregards the initial textual case description when additional image prompts are introduced, necessitating its repetition for accurate responses. As the context window of the model running in the web application is unknown, as well as the image resolution and embedding size, definitive conclusions regarding the cause of this forgetfulness were not possible. Nevertheless, the model's consistent ability to recognize and recall images from the initial prompt (e.g., axial FLAIR images) suggests that running out of context length is an unlikely explanation. The models were accessed between November 6th, 2023, and November 17th, 2024. Our team utilized an API account, which allowed us to employ the models programmatically and ensure a consistent environment for each test. This access level is crucial as it provides a stable and repeatable interaction with the models, unlike what might be experienced with fluctuating conditions of regular account usage. The ground truth for each case was established based on the final diagnosis as stated in the RSNA case entries.
  \newline
  \textit{Reproduced with permission from the Radiological Society of North America. Link to the displayed case: https://cases.rsna.org/take-quiz/07c4b917-80fb-43c0-8b3b-59a0d8ceb203 (accessed 14th January 2026).}}
  \label{fig:fig1}
\end{figure}

We sourced 207 cases with 1312 images from the Radiological Society of North America (RSNA) Case Collection, accessible at https://cases.rsna.org/, to obtain a representative sample from each subspecialty. Each model's evaluation comprised two tasks repeated three times to assess consistency and reliability and to increase the robustness of our findings. First, the model was asked to recognize the diagnosis and offer two differential diagnoses; second, it was asked to select the appropriate answer from a set of multiple-choice questions. The mean accuracy across the three iterations was calculated, with bootstrapped 95\% confidence intervals (CI) to provide a measure of precision. The McNemar test was used to test for statistical significance, with a p-value of <.001 considered significant.
For the evaluation of GPT-4V, we used a ‘chain-of-thought’ style prompting approach. The 'chain-of-thought' prompting technique involves guiding the AI model through a step-by-step reasoning process, much like how a human would approach a problem-solving task. The model was first instructed to describe the images and then to make a diagnostic assessment. An illustrative dialog is shown in Figure \ref{fig:fig1}. For comparison, the GPT-4 model was assessed using case descriptions only.
 
\section{Results} 
\begin{figure}
  \centering
  \includegraphics[scale=0.35]{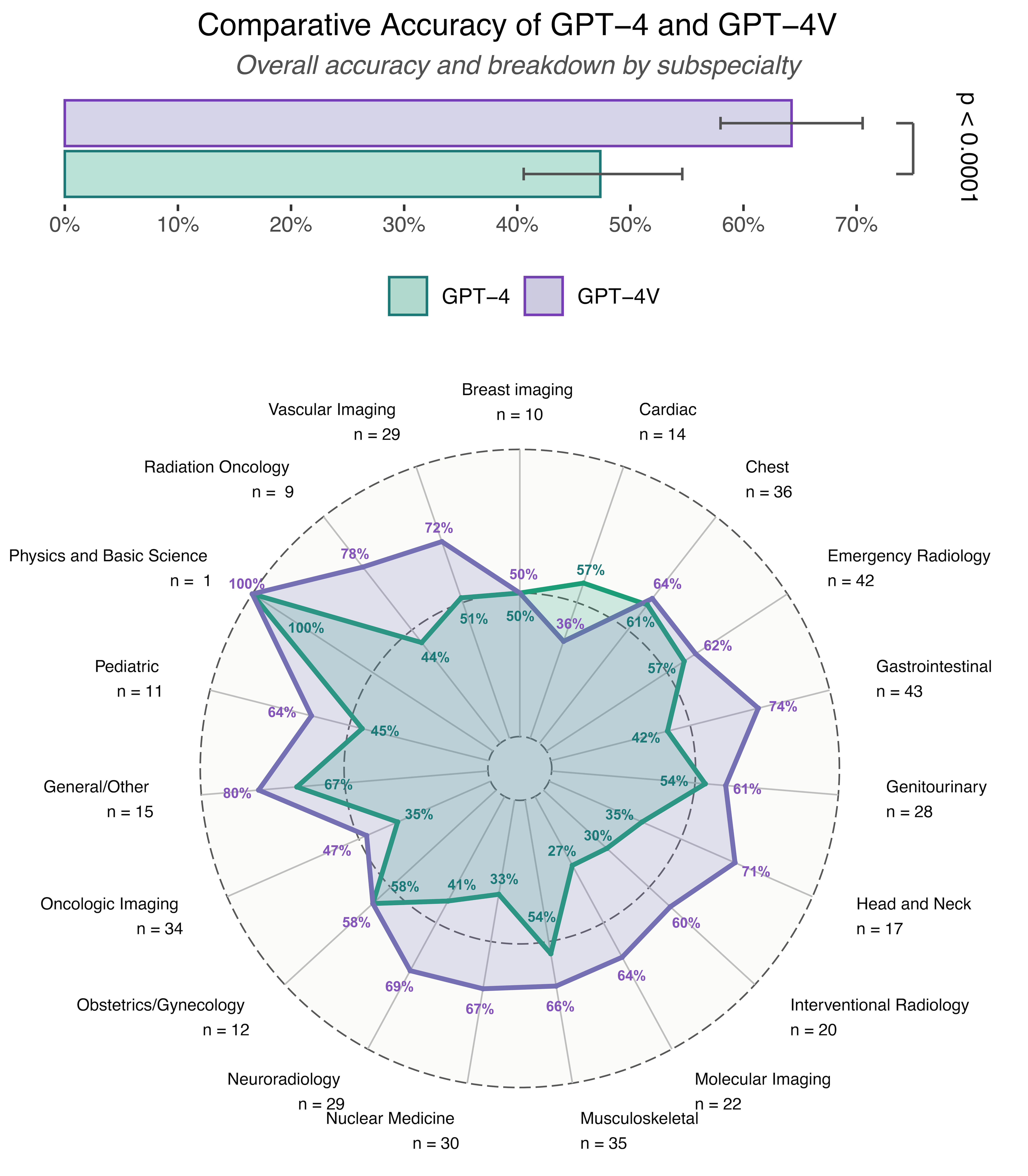}
  \caption{Comparison of GPT-4 and GPT-4V in various radiology subspecialties. Many cases spanned multiple subspecialties, and some subspecialties had very few cases. Individual cases ranged from 2 to 30 images, and the overall accuracy across all subspecialties, as shown in the bar plot, showed a significantly better performance of GPT-4 Vision over GPT-4. Error bars represent the 95\% confidence interval. The radar plot shows the accuracy of GPT-4 (green line) and GPT-4V (purple line) across different radiology subspecialties. Each axis represents a specific radiology subspecialty, with the value indicating the accuracy of the model in that domain. Both models show varying levels of performance across subspecialties, with GPT-4V consistently performing better than GPT-4, except for cardiac imaging (n=14; GPT-4V: 36\% versus GPT-4: 57\%). This finding warrants cautious interpretation due to the relatively small number of cases in this subgroup. For physics and basic science (n=1), breast imaging (n=10), and obstetrics/gynecology (n=12), GPT-4V and GPT4 showed on-par performance (100\%, 50\% and 58\%, respectively). The small sample sizes in these categories limit the statistical robustness of our findings. Therefore, these results should primarily be viewed as indicative trends rather than definitive conclusions about the models' performance in these specific areas. For the category physics and basic science, no assessment is possible, as only one case was included in this category.}
  \label{fig:fig2}
\end{figure}

GPT-4 accurately identified the primary diagnosis in 18\% (95\% CI: 12\%-25\%) of cases (first task). When including differential diagnoses, this accuracy increased to 28\% (95\% CI: 22\%-33\%). In contrast, GPT-4V achieved a 27\% (95\% CI: 21\%-34\%) accuracy rate for primary diagnosis, which increased to 35\% (95\% CI: 29\%-40\%) when differential diagnoses were accounted for. After being presented with multiple choice questions, including clinical history and presentation (second task), GPT-4 achieved an accuracy of 47\% (95\% CI: 42\%-56\%). Again, GPT-4V demonstrated a higher accuracy of 64\% (95\% CI: 59\%-72\%). The observed difference in performance was statistically significant with a p-value < 0.0001. Across 15 subspecialties, GPT-4V outperformed GPT-4, with the sole exception being in cardiac imaging. Figure \ref{fig:fig2} summarizes the accuracies across all subspecialties.
 
\section{Discussion}
Our study suggests that GPT-4V, a multimodal AI model, outperforms GPT-4 in handling complex radiology questions, hinting at its ability to identify pathological features in medical images. The complexity of the RSNA cases, tailored for professional radiologists, sets a high standard for accuracy and highlights GPT-4V's potential in advanced medical contexts.
While the performance of the GPT-4V is promising, it should be viewed as a potential adjunct tool that may in the future augment, but not replace, the nuanced analysis provided by trained medical professionals. 
This study extends the preliminary work on GPT-4V's medical image analysis capabilities, as explored by Yang et al. \cite{Yang2023-wc}, into more complex scenarios and a broader case range. The development of other multimodal models like Med-Flamingo \cite{Moor2023-un}, geared towards medical applications, also illustrates the growing interest in this area.
A key challenge in this field is the lack of robust, specialized medical datasets. The exclusive access of RSNA members to the cases used in our study reduces the risk of data contamination, making it unlikely that these cases were included in GPT-4V’s training data.
A limitation of our study is the selective nature of the images presented to the models, chosen specifically to highlight certain pathologies. This approach does not mirror the comprehensive analysis required in clinical practice, where a full examination of all image slices is standard. This task is currently more effectively managed by specialized deep learning models rather than GPT-4V.
While GPT-4V demonstrates adaptability to specialized domains, the variability in sample sizes across subspecialties calls for further, more extensive research. Our results, which show the advantage of GPT-4V over GPT-4, underscore the need for a broader exploration with large datasets to fully determine their role in radiology.

\bibliographystyle{unsrt}  
\bibliography{references}

@ARTICLE{Yang2023-wc,
  title    = "The Dawn of {LMMs}: Preliminary Explorations with {GPT-4V(ision})",
  author   = "Yang, Zhengyuan and Li, Linjie and Lin, Kevin and Wang, Jianfeng
              and Lin, Chung-Ching and Liu, Zicheng and Wang, Lijuan",
  month    =  sep,
  year     =  2023,
  eprint   = "2309.17421"
}

@ARTICLE{Moor2023-un,
  title    = "{Med-Flamingo}: a Multimodal Medical Few-shot Learner",
  author   = "Moor, Michael and Huang, Qian and Wu, Shirley and Yasunaga,
              Michihiro and Zakka, Cyril and Dalmia, Yash and Reis, Eduardo
              Pontes and Rajpurkar, Pranav and Leskovec, Jure",
  month    =  jul,
  year     =  2023,
  eprint   = "2307.15189"
}

@ARTICLE{Nori2023-ly,
  title    = "Capabilities of {GPT-4} on Medical Challenge Problems",
  author   = "Nori, Harsha and King, Nicholas and McKinney, Scott Mayer and
              Carignan, Dean and Horvitz, Eric",
  month    =  mar,
  year     =  2023,
  eprint   = "2303.13375"
}

@ARTICLE{Bhayana2023-hm,
  title     = "Performance of {ChatGPT} on a Radiology Board-style Examination:
               Insights into Current Strengths and Limitations",
  author    = "Bhayana, Rajesh and Krishna, Satheesh and Bleakney, Robert R",
  journal   = "Radiology",
  publisher = "Radiological Society of North America",
  month     =  may,
  year      =  2023,
  language  = "en"
}

@ARTICLE{Adams2023-dn,
  title    = "Leveraging {GPT-4} for Post Hoc Transformation of Free-text
              Radiology Reports into Structured Reporting: A Multilingual
              Feasibility Study",
  author   = "Adams, Lisa C and Truhn, Daniel and Busch, Felix and Kader, Avan
              and Niehues, Stefan M and Makowski, Marcus R and Bressem, Keno K",
  journal  = "Radiology",
  volume   =  307,
  number   =  4,
  pages    = "e230725",
  month    =  may,
  year     =  2023,
  language = "en"
}

@MISC{noauthor_undated-vh,
  title        = "{GPT-4V(ision}) system card",
  howpublished = "\url{https://openai.com/research/gpt-4v-system-card}",
  note         = "Accessed: 2023-10-14",
  language     = "en"
}

\end{document}